\documentclass{article}
\usepackage[top=1.5in, bottom=1.5in, left=1.2in, right=1.2in]{geometry}
\usepackage{authblk}
\usepackage{amssymb, amsmath,framed}
\usepackage{graphicx}
\usepackage[round]{natbib}

\usepackage{bm}
\usepackage[colorlinks=true,linkcolor=blue,urlcolor=blue,citecolor=blue,anchorcolor=blue]{hyperref}
\expandafter\ifx\csname package@font\endcsname\relax\else
 \expandafter\expandafter
 \expandafter\usepackage
 \expandafter\expandafter
 \expandafter{\csname package@font\endcsname}
\fi
\def\bq{\begin{equation}}
\def\eq{\end{equation}}
\def\bqy{\begin{eqnarray}}
\def\eqy{\end{eqnarray}}

\makeatother

\begin{document}

\title{Photosynthesis on Exoplanets and Exomoons from Reflected Light}

\author{Manasvi Lingam \thanks{Electronic address: \texttt{manasvi.lingam@cfa.harvard.edu}}}
\affil{Institute for Theory and Computation, Harvard University, 60 Garden St, Cambridge MA 02138, USA}
\affil{Department of Aerospace, Physics and Space Sciences, Florida Institute of Technology, 150 W University Blvd, Melbourne FL 32901, USA}

\author{Abraham Loeb \thanks{Electronic address: \texttt{aloeb@cfa.harvard.edu}}}
\affil{Institute for Theory and Computation, Harvard University, 60 Garden St, Cambridge MA 02138, USA}

\date{}

\maketitle

\begin{abstract}
Photosynthesis offers a convenient means of sustaining biospheres. We quantify the constraints for photosynthesis to be functional on the permanent nightside of tidally locked rocky exoplanets via reflected light from their exomoons. We show that the exomoons must be at least half the size of Earth's moon in order for conventional oxygenic photosynthesis to operate. This scenario of photosynthesis is unlikely for exoplanets around late-type M-dwarfs due to the low likelihood of large exomoons and their orbital instability over long timescales. Subsequently, we investigate the prospects for photosynthesis on habitable exomoons via reflected light from the giant planets that they orbit. Our analysis indicates that such photosynthetic biospheres are potentially sustainable on these moons except those around late-type M-dwarfs. We conclude our analysis by delineating certain physiological and biochemical features of photosynthesis and other carbon fixation pathways, and the likelihood of their evolution on habitable planets and moons.
\end{abstract}

\section{Introduction} \label{SecIntro}
The overwhelming majority of Earth's biomass is dependent, either directly or indirectly, on photosynthesis for its maintenance and growth \citep{BPM18}. This fact is not particularly surprising given that solar radiation constitutes the most dominant free energy source on Earth \citep{Deam97}. Photosynthesis evolved early in our planet's history - perhaps as early as $\gtrsim 3.7$ Ga - and the advent of oxygenic photosynthesis led to a major transformation of Earth's geochemical and biological landscape \citep{Knoll15}. The existence of photosynthesis is not only important from the standpoint of sustaining complex biospheres but also as a flag enabling the detection of biosignatures via remote sensing. As oxygenic photosynthesis yields molecular oxygen as a product, much effort has been devoted to modeling the feasibility of detecting biogenic O$_2$ via spectroscopy \citep{MRA18}. Another notable consequence of photosynthesis is the manifestation of the ``vegetation red edge'' that may be discernible through spectral observations \citep{STSF}.

For these reasons, a great deal of effort has been devoted to studying the prospects for photosynthesis on other planets and moons. For instance, several studies suggest that the net primary productivity of M-dwarf exoplanets is lower than the Earth \citep{Pol79,RLR18} and that planets orbiting late-type M-dwarfs might not build up sufficient atmospheric O$_2$ despite the presence of photosynthetic lifeforms \citep{LCP18,LiLo19}. It is, however, important to move away from the conventional paradigm of evaluating photosynthesis on an Earth-like planet orbiting a solar-type star and consider other possibilities. For instance, other studies of photosynthesis have explored environments as diverse as water worlds \citep{ML19}, planets in binary and multiple star systems \citep{ORC12,FMC15}, planets orbiting brown dwarfs \citep{RaD13}, brown dwarf atmospheres \citep{LL19}, near black smokers \citep{BOL05}, and artificial lights \citep{RC06}.

In this paper, we investigate two distinct scenarios. In the first, we consider potentially tidally locked exoplanets with a permanent nightside, on which photosynthesis is assumed to take place via reflected light from an exomoon orbiting the planet. In the second case, we address photosynthesis on the nightside of an Earth-like habitable exomoon via light reflected from a giant planet around which the moon orbits. Both of these scenarios have been explored in \citet{RC06} and \citet{CRKL09}. Our work differs from these two studies in the following respects. First, we quantify the feasibility of photosynthesis not only for Sun-like stars but also for K- and M-dwarfs. Second, we carry out a systematic analysis of the allowed planet-star separations while taking the sizes of the planet and moon as well as other constraints on habitability into account.

The outline of the paper is as follows. In Sec. \ref{SecMath}, we describe the mathematical setup to determine the photon fluxes received via reflected light. Next, we study the prospects for photosynthesis on Earth-like planets and moons, while taking habitability constraints into account, in Sec. \ref{SecPPM}. We follow this up with a discussion of the basic physiology and biochemistry of photosynthesis, its relation to other carbon fixation pathways, and the prospects for its evolution on other worlds in Sec. \ref{SecBCP}. We end with a summary of our central results in Sec. \ref{SecConc}. 

\section{Mathematical set-up}\label{SecMath}
There are two distinct cases that we shall investigate, but they can be tackled using the same formalism. In the first, reflected light from an exomoon illuminates the nightside of a tidally locked rocky planet. In the second, reflected light from a Jovian planet illuminates a large and habitable exomoon orbiting it. In both instances, we will refer to the object from which light is reflected as the ``primary'' and the object on which the reflected light is incident as the ``secondary''. We use the subscripts `P' and `S' to denote the quantities associated with the primary and secondary objects, while the subscript `$\star$' labels stellar parameters. 

In the subsequent analysis, we define photosynthetically active radiation (PAR) as having minimum and maximum wavelengths of $\lambda_\mathrm{min} = 350$ nm and $\lambda_\mathrm{max} = 750$ nm, respectively \citep{CB11,NMS18}. We have deliberately opted to choose a conservative choice based on the limits for oxygenic photoautotrophs on Earth. In theory, it is conceivable that the maximum wavelength for PAR could extend into the near-infrared (near-IR) if multiple photons are utilized per electron transfer, as opposed to Earth-based oxygenic photosynthesis with its two photons per electron mechanism \citep{WoRa02}. The minimum wavelength for PAR is also not well constrained, but theoretical models suggest that the choice of $\sim 350$ nm might be fairly reasonable for photosystems akin to those found on Earth \citep{CoAi}. Note that, for the most part, we do not take more exotic versions of photosynthesis such as ``chlorinic'' \citep{Haas10} or ``hydrogenic'' photosynthesis \citep{BSZ14} into account in this paper.

For the sake of simplicity, we assume that the planet under consideration is orbiting the star in a roughly circular orbit and that it receives approximately the same stellar flux as the Earth, thus placing it either within or close to the circumstellar habitable zone \citep{KWR93,Ram18}. In this event, its orbital radius is
\begin{equation}
    a = 1\,\mathrm{AU}\,\left(\frac{L_\star}{L_\odot}\right)^{1/2}  = 1\,\mathrm{AU}\,\left(\frac{R_\star}{R_\odot}\right)\left(\frac{T_\star}{T_\odot}\right)^2,
\end{equation}
where the last equality follows from the black body relation for the stellar luminosity. Therefore, the photon flux received by the primary is given by
\begin{equation}
    \Phi_P \approx \frac{\dot{N}_\star}{4\pi a^2},
\end{equation}
where the number of photons (comprising PAR) emitted by the star per unit time ($\dot{N}_\star$) is
\begin{equation}\label{NstarBB}
\dot{N}_\star = 4 \pi R_\star^2 \int_{\lambda_\mathrm{min}}^{\lambda_\mathrm{max}} \frac{2c}{\lambda^4}\left[\exp\left(\frac{h c}{\lambda k_B T_\star}\right)-1\right]^{-1}\,d\lambda,
\end{equation}
assuming a black body spectrum. It is fairly reasonable to model stars as black bodies since the contributions from flares and other stellar processes are not likely to contribute significantly to the PAR flux in most instances \citep{LiLo19}. What we wish to determine, however, is the maximum PAR flux incident on the secondary object ($\Phi_S$). It can be estimated from $\Phi_P$ using the following formula:
\begin{equation}\label{PhiSDef}
    \Phi_S \approx \frac{R_P^2 A_P \Phi_P}{2d^2},
\end{equation}
where $R_P$ and $A_P$ are the radius and albedo (in the PAR range) of the primary, whereas $d$ denotes the orbital radius of the moon around the planet assuming an approximately circular orbit. In deriving the above formula, we have presumed that the atmosphere of the secondary object (i.e., the habitable world under question) is similar to the Earth insofar as its optical depth for PAR is concerned; in other words, the atmosphere is assumed to be optically thin to incoming PAR from the primary object. 

As there are several free parameters, we will introduce a few assumptions to simplify our analysis. As noted earlier, we shall work with the conservative choice of $\lambda_\mathrm{min} = 350$ and $\lambda_\mathrm{max} = 750$ nm as these limits are well-documented on Earth. In actuality, the maximum wavelength for PAR may extend to $\sim 1$ $\mu$m for planets orbiting M-dwarfs \citep{WoRa02,HDJ99,KST07} and $\sim 2$-$3$ $\mu$m for atmospheric habitable zones in brown dwarfs \citep{LL19}. Second, we specify a fiducial value of $A_P = 0.2$ because it is only a factor of $\lesssim 2$ removed from the visual albedos of most Solar system bodies. 

After employing the above relations and simplifying (\ref{PhiSDef}), we end up with
\begin{eqnarray}\label{PhiSFin}
&& \Phi_S \approx 2.3 \times 10^{16}\,\mathrm{m^{-2}\,s^{-1}}\,\left(\frac{A_P}{0.2}\right)\left(\frac{R_P}{R_\oplus}\right)^2  \nonumber \\
&& \hspace{0.3in} \times \left(\frac{d}{60\,R_\oplus}\right)^{-2} \left(\frac{T_\star}{T_\odot}\right)^{-1} \mathcal{F}(T_\star), 
 \end{eqnarray}
where our normalization for $d$ is based on the current Earth-Moon separation, $T_\odot$ is the black body temperature of the Sun, and the function $\mathcal{F}$ is defined as
\begin{equation}
\mathcal{F}(T_\star) \approx \int_{x_1(T_\star)}^{x_2(T_\star)} \frac{x'^2\,dx'}{\exp\left(x'\right) - 1},    
\end{equation}
with $x_1 \approx 3.32 \left(T_\star/T_\odot\right)^{-1}$ and $x_2 \approx 7.12 \left(T_\star/T_\odot\right)^{-1}$. If we consider the Earth-Moon system, upon specifying $A_P = 0.12$ and $R_P = 0.27 R_\oplus$, we obtain $\Phi_S \approx 6.4 \times 10^{14}$ photons m$^{-2}$ s$^{-1}$. This result is in good agreement with empirical data concerning PAR fluxes arising from the full Moon; estimates for the latter range from $\sim 3$-$70 \times 10^{14}$ photons m$^{-2}$ s$^{-1}$ \citep{GF02,JAW06,CBR08,CRKL09}.

In order for Earth-like photosynthesis to function, a minimum photon flux is necessary. This lower limit can be determined from physicochemical considerations and has a value of $\Phi_c \approx 1.2 \times 10^{16}$ m$^{-2}$ s$^{-1}$ for photosynthetic organisms on Earth \citep{RKB00}. Thus, by imposing the fact that $\Phi_S \gtrsim \Phi_c$, we arrive at the following inequality:
\begin{equation}\label{InEqu}
\left(\frac{A_P}{0.2}\right)\left(\frac{R_P}{R_\oplus}\right)^2 \left(\frac{d}{60\,R_\oplus}\right)^{-2} \left(\frac{T_\star}{T_\odot}\right)^{-1} \mathcal{F}(T_\star) \gtrsim 0.5
\end{equation}
The left-hand-side of the above equation has a dependence on four different parameters. Henceforth, we will hold $A_P$ constant for the reasons elucidated earlier and investigate the dependence on the other three variables.

\section{Photosynthesis on planets and moons}\label{SecPPM}
We will now tackle the two distinct cases that were outlined in Secs. \ref{SecIntro} and \ref{SecMath}. 

\subsection{Photosynthesis on Earth-like planets}\label{SSecELP}

\begin{figure}
\includegraphics[width=7.5cm]{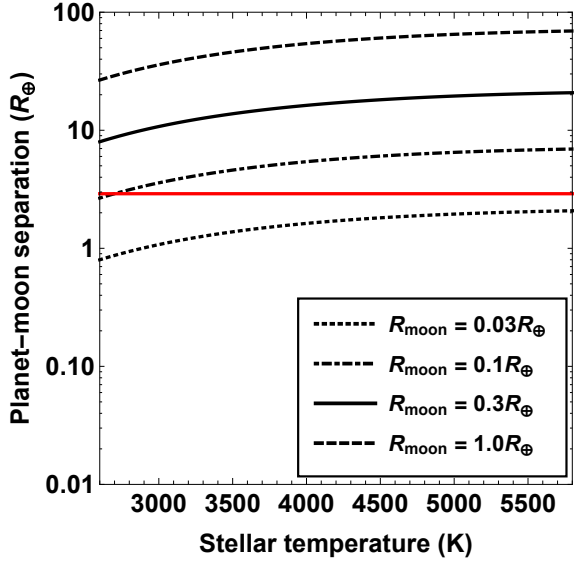} \\
\caption{The maximum separation ($d$) between the planet and the moon (in units of $R_\oplus$) for photosynthesis to occur on the nightside of the planet at full moon, as a function of the stellar temperature (in K). The various curves correspond to $d$ for different exomoon sizes. The horizontal red line corresponds to the Roche limit for an Earth-analog assuming that the exomoon's composition is similar to that of Earth's moon.}
\label{FigMoonP}
\end{figure}

\begin{figure}
\includegraphics[width=7.5cm]{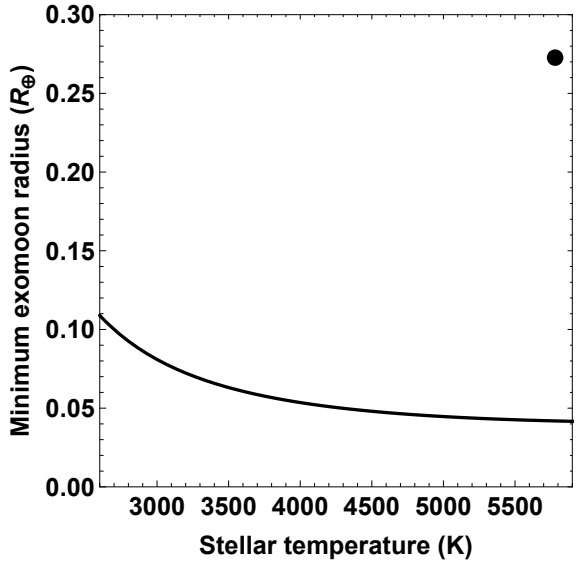} \\
\caption{The minimum moon radius (in $R_\oplus$) required in order to enable photosynthesis to occur on the nightside of a tidally locked planet, as a function of the stellar temperature (in K). The parameters for Earth's moon (the black dot) are shown for reference.}
\label{FigMSize}
\end{figure}

This scenario corresponds to a tidally locked exoplanet orbiting a star, which is expected to be ubiquitous for planets in the habitable zone of dwarf stars \citep{Barn17}. It is conceivable that some of the best-known planets discovered in recent times such as Proxima b \citep{AA16} and the seven planets around TRAPPIST-1 \citep{GTD17} might belong to this category. As the nightside would always face away from the star, it cannot support photosynthesis on its own because it does not receive stellar radiation. However, the existence of an exomoon can, perhaps, enable photosynthesis at full moon on the planetary nightside provided that (\ref{InEqu}) is satisfied. 

Fig. \ref{FigMoonP} shows the maximum separation between the planet and the moon ($d$) that still permits photosynthesis to occur on the nightside at full moon as a function of the stellar temperature for different exomoon sizes.\footnote{We have chosen to truncate the stellar temperature in the plots at $T_\star \approx T_\odot$, as it is unlikely for exoplanets to be tidally locked around more massive stars over Gyr timescales, except under special circumstances \citep{Barn17}.} As the moon size gets smaller, $d$ also decreases along expected lines. When the stellar temperature is lowered, fewer PAR photons are received, causing $d$ to decrease in order to compensate for the reduction in PAR flux. We have also plotted the Roche limit ($d_L$) for an Earth-like planet under the assumption that its exomoon has a mean density comparable to the Moon; for fluid satellites, $d_L$ is expressible as
\begin{equation}\label{RoLim}
   d_L \approx 2.46 R_\mathrm{planet} \left(\frac{\rho_\mathrm{planet}}{\rho_\mathrm{moon}}\right)^{1/3}, 
\end{equation}
where $R_\mathrm{planet}$ is the planet's radius, while $\rho_\mathrm{planet}$ and $\rho_\mathrm{moon}$ are the densities of the planet and its moon, respectively \citep{MD99}. The significance of the Roche limit stems from the fact that $d < d_L$ would lead to disruption of the exomoon due to tidal forces exerted by the planet. 

If we substitute $d = d_L$ in (\ref{InEqu}), we can determine the lower bound on the radius of the exomoon as a function of the stellar temperature. The resulting criterion has been plotted in Fig. \ref{FigMSize}. This figure implies that the minimum exomoon radius must be approximately half the radius of the Earth's moon. As no exomoons have been conclusively identified so far,\footnote{The evidence for a Neptune-sized exomoon orbiting Kepler-1625b \citep{TK18} is ambiguous, and other interpretations have been proposed \citep{KLB19}.} the frequency of large exomoons as a function of the star spectral type remains unknown. However, theoretical considerations suggest that compact exoplanetary systems around low-mass star (e.g., TRAPPIST-1) have a low likelihood of hosting exomoons \citep{Kane17}.

There is another vital issue that must be taken into account. If the exomoon's orbit is not stable, any photosynthesis driven by it will be transient in nature. Hence, it is important for the exomoon to be able to survive over long timescales without escaping the planet or being disrupted. The issue of the orbital stability of exomoons is complex because it is sensitive to the initial spin period of the planet, the tidal dissipation factor of the planet, the mass of the exomoon, the initial moon-planet and planet-star separation, the orientation of their orbital planes, and the spectral type of the host star among other factors. 

\citet{SB14} carried out numerical simulations and found that stars with stellar mass $M_\star < 0.4 M_\odot$ were unlikely to host exomoons over Gyr timescales for a wide range of bulk compositions for the planet-moon system. On the other hand, numerical results from \citet{Piro18} indicate that stars with $M_\star < 0.5 M_\odot$ might be able to retain their moons over timescales of $\sim 10^9$ yrs if the planet was initially situated outside the habitable zone before potentially migrating inwards. This inward migration could have occurred for the planets of the TRAPPIST-1 system \citep{UDH18} and other planetary systems detected by the \emph{Kepler} mission \citep{WF15}.

\subsection{Photosynthesis on Earth-like moons}

\begin{figure}
\includegraphics[width=7.5cm]{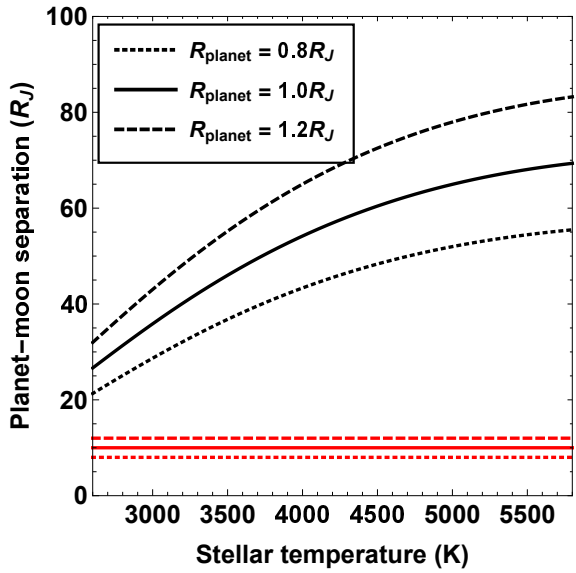} \\
\caption{The maximum separation ($d$ in units of $R_J$) between a giant planet and an Earth-like moon for photosynthesis on the moon at night (via reflected light from the planet), as a function of the stellar temperature (in K). The various black curves correspond to $d$ for different radii of the giant planet. The red curves (for the corresponding planetary sizes) depict the cutoff distances for $d$ that must be exceeded in order to ensure that the moon is habitable.}
\label{FigELMoon}
\end{figure}

The second scenario we consider is a large exomoon with an Earth-like atmosphere (albeit not necessarily the same size) orbiting a gas giant planet in the habitable zone \citep{WKW97,HWK14}. In this setting, starlight reflected from the giant planet would illuminate the moon during its night and enable photosynthesis; the relevant geometry for this case has been illustrated in \citet{CRKL09}.

We can estimate the constraints on the planet-moon separation by making use of (\ref{InEqu}) and carrying out an analysis along the lines of Sec. \ref{SSecELP}. However, it is important to appreciate a couple of distinctions. Recall that, as per our notation, $R_P$ now refers to the radius of the gas giant, which we shall measure in units of Jupiter's radius ($R_J$). Second, by using (\ref{RoLim}), we find that the Roche limit is $d_L \approx 1.53 R_\mathrm{planet}$ after supposing that the densities of the giant planet and the habitable exomoon are similar to that of Jupiter and Earth, respectively.

However, this is not the only constraint on the planet-moon separation ($d$). Assessing the habitable zone for an exomoon is a complex endeavor because it depends not only on the properties of the classical circumstellar habitable zone (e.g., stellar flux) but also the eccentricity of the moon's orbit, its inclination to the ecliptic, its rheology, the mass of the giant planet, and the value of $d$ to name a few \citep{HWK14,DT15,FD16,DHT17}. In view of this complexity, it is difficult to identify a realistic lower bound on $d$. However, when the stellar insolation received by the planet-moon system is similar to that incident on the Earth, a cutoff of $d_\mathrm{min} \approx 10 R_\mathrm{planet}$ appears to be reasonable \citep{HB15,ZAH17}. When $d < d_\mathrm{min}$, the planet is susceptible to a runaway greenhouse effect for $\mathcal{O}\left(10^8\right)$ yr, and could therefore end up losing much of its water inventory during this period \citep{HB15}.

The maximum planet-moon separation that permits photosynthesis at night on the exomoon by way of reflected light from the giant planet is plotted in Fig. \ref{FigELMoon}. At all stellar temperatures, we find that $d > d_\mathrm{min}$. Hence, it would seem as though there exist regions of parameter space where the exomoon is situated sufficiently far from the planet so as to remain habitable while simultaneously able to receive enough PAR to power photosynthesis via reflected light. 

However, there is another factor that needs to be taken into consideration. As the habitable zones of low-mass stars are located at close-in distances, any exomoons in this region are subject to strong tidal torques from the star. Numerical models indicate that exomoons in habitable zones around stars with $M_\star \lesssim 0.2 M_\odot$ are unlikely to be habitable because of stellar perturbations, and even those around stars with $0.2 M_\odot < M_\star \lesssim 0.5 M_\odot$ may experience considerable stellar gravitational effects \citep{Hell12,ZAH17}.

\section{Looking beyond conventional photosynthesis}\label{SecBCP}
Until now, we have primarily focused on investigating the constraints that permit ``conventional'' oxygenic photosynthesis to function on planets and moons on the nightside. We will briefly delve into other possibilities herein and explore the ensuing ramifications for biospheres.

\subsection{The basis of photosynthesis}
The photosynthetic machinery inherent to organisms on Earth is intricate and characterized by its complex biochemistry and physiology. Hence, it is not immediately obvious \emph{a priori} as to which features found on Earth-based photoautotrophs would also be manifested on other habitable worlds. For this reason, we will focus on highlighting only a few generic features of photosynthesis, with an emphasis on oxygenic photosynthesis, which might exhibit some degree of universality. Comprehensive reviews of this subject can be found in \citet{HB11}, \citet{Bla14}, \citet{NJ15} and \citet{FHJ16}.

In its most basic form, the net reaction of photosynthesis is expressible as follows:
\begin{equation}
    \mathrm{CO_2} + 2 \mathrm{H_2X} \, \xrightarrow[\text{pigments}]{h \nu}\, \mathrm{CH_2O} + \mathrm{H_2O} + 2 \mathrm{X}.
\end{equation}
In the above equation, H$_2$X denotes the reducing agent (i.e., electron donor) that undergoes biochemical oxidation to yield electrons that are utilized in subsequent biochemical reactions. Examples of reducing agents used in photosynthesis include H$_2$S, H$_2$, and H$_2$O; for the latter, note that O$_2$ is the metabolic waste product. The product $\mathrm{CH_2O}$ essentially represents a reduced carbon compound (e.g., sugar) where the energy is stored. The net reaction is endergonic in nature, owing to which the input of light energy (exemplified by $h \nu$) is necessary.

In a recent review, \citet{SKP18} posited that three basic stages ought to be operational in a generic photosynthetic apparatus (reaction centre). The photosynthetic reactions are initiated via the photoelectric effect and relies on the absorption of photons by a suitable pigment to produce electrons in an excited state. Given sufficient energies, the electrons are ejected from the molecule, thus leaving behind an electron hole. The ejected electron must be replaced, which can happen either through cyclical or non-cyclical electron transfer mechanisms. In the case of the latter, the ejected electron is replaced when the biomolecule(s) in the photosystem under question oxidizes the reducing agent (H$_2$X) and yields the metabolic product X.  The energy inherent to the ejected electron is used for two purposes: the oxidation of the reducing agent and the synthesis of reduced carbon compounds (which act as repositories for the captured energy) via redox reactions.  

An important point to recognize here is that the photon energy is not directly used for photolysis of the reductant. Instead, as noted above, the oxidation of the reducing agent requires the biomolecule(s) comprising the photosystem to be more oxidizing than the former. Bearing this fact in mind, we turn our attention to the potential reductants. The redox potentials for H$_2$/2H$^+$ and H$_2$S/S$^0$ are $-0.42$ V and $-0.24$ V, respectively, at neutral pH \citep{HB11}. In contrast, the redox potential for the H$_2$O/O$_2$ pair is $+ 0.815$ V \citep{HB11}. In other words, it is relatively easier to extract electrons from strong reductants such as H$_2$ and H$_2$S. Hence, it is not surprising that microbes reliant on these reductants possess comparatively simpler photosynthetic machinery, i.e., they have only a single photosystem (PSI or PSII). It is commonly supposed that the anoxygenic photosynthesis (with its single photosystem) evolved on Earth earlier than its oxygenic counterpart (which has two photosystems), but the evidence favouring this hypothesis has been challenged as of late \citep{Card19}.

For the time being, let us adopt the notion that anoxygenic photosynthesis would have evolved more readily on other worlds because the presence of stronger reductants (e.g., H$_2$S) would impose less stringent constraints on the oxidizing biomolecule(s) in the photosystem. However, at this juncture, we encounter a potential bottleneck imposed by geology, namely, the available fluxes of these reductants. On Earth, the geological fluxes of electron donors for photosynthesis were probably limited, which in turn may have yielded a net primary productivity (NPP) that was $\sim 3$ orders of magnitude smaller than the present-day value \citep{WRF19,WaSh19}. Once water could be utilized as an electron donor, the bottleneck on NPP was possibly eliminated; other factors such as nutrients (e.g., phosphorus) would have limited the NPP instead. 

Therefore, unless other worlds have a much higher inventory of volcanogenic reducing agents, it is likely that higher NPP is typically achievable by the use of water as an electron donor. However, as mentioned earlier, the redox potential for the water-oxygen pair is very high with respect to other reducing agents commonly employed in anoxygenic photosynthesis. Hence, several authors have suggested that intermediate reducing agents such as Fe$^{2+}$ and Mn$^{2+}$, especially the latter, may have served as transitional electron donors \citep{FHJ16,MRA18}; for instance, the redox potential for the Fe$^{2+}$/Fe$^{3+}$ pair at neutral pH is $\sim 0.2$ V \citep{HB11}. The oxidation of water in photoautotrophs on Earth is facilitated by the water-oxidizing complex (WOC) situated in photosystem II (PSII). The core of the WOC is a manganese cluster (Mn$_4$CaO$_5$), whose oxidation states, thermodynamics and kinetics are described in \citet{VAC13}, \citet{WNAS} and \citet{NJ15}. 

All oxygenic photoautotrophs on Earth rely upon the manganese cluster (in the WOC) for the purpose of evolving molecular oxygen. Hence, it is natural to wonder whether other variants of the WOC are feasible. Although no such examples appear to exist in photoautotrophs, several alternatives have been artificially designed in the laboratory. Some of the alternatives to manganese in the WOCs include copper (Cu), nickel (Ni), ruthenium (Ru) and iridium (Ir); reviews of this rapidly growing subject can be found in \citet{BCB15}, \citet{LGM17}, \citet{SHQ17} and \citet{ZS19}. Molecular catalysts synthesized using these elements enable the ``splitting'' of water to yield molecular oxygen as follows:
\begin{equation}
    2 \mathrm{H_2O} \rightarrow \mathrm{O_2} + 4 \mathrm{H^+} + 4 \mathrm{e^-}
\end{equation}
In principle, therefore, it is conceivable that WOCs reliant on the likes of copper or nickel clusters instead of manganese might evolve on other planets and moons. 

When it comes to light-harvesting pigments, it is important to distinguish between the antenna pigments that absorb photons (of different wavelengths) and transmit them to the reaction centre (RC) pigment, which can donate electrons by absorbing photons of a particular wavelength and undergoing excitation across the band gap \citep{KSG07}. The colour and biosignatures produced by photosynthetic organisms are dependent not only on the RC pigment but also on the antenna pigments. It is not easy to determine over what wavelengths pigments will optimally absorb radiation because it is governed by the oxidation state of the pigment macrocycle as well as the functional groups and proteins surrounding the macrocycle. The peak absorption wavelengths for light-harvesting pigments range from $\sim 0.7$-$1.0$ $\mu$m for bacteriochlorophylls to $\sim 0.4$ $\mu$m and $\sim 0.7$ $\mu$m for chlorophylls \citep{SKP18}. Another notable light-harvesting pigment, bacteriorhodopsin, exhibits a peak of $\sim 0.6$ $\mu$m \citep{DSS18}. 

In spite of the fact that no convincing alternatives to tetrapyrrole-based pigments (e.g., chlorophylls) have been identified thus far, it is difficult to estimate what factors will govern the peak absorbance of pigments on other worlds. This issue was explored by \citet{KSG07} wherein it was suggested that the peak absorbance of exo-pigments might occur near: (a) the wavelength associated with the maximum value of the incident photon flux density or (b) the longest wavelength that permits the resonance transfer of excitation energy and energy funnelling in antenna and RC pigments. If we focus on (a), it is apparent that the peak absorbance will be shifted toward longer wavelengths on M-dwarfs as the peak photon flux density of these stars occurs in the near-infrared. There is also an extra complication introduced by atmospheric transmission, which will depend on the chemical composition and bulk properties of the atmosphere. As the latter is empirically unknown for habitable exoplanets (or exomoons) at this stage, we will restrict ourselves to Earth-like worlds. 

\subsection{Alternatives to conventional oxygenic photosynthesis}

\begin{figure}
\includegraphics[width=7.5cm]{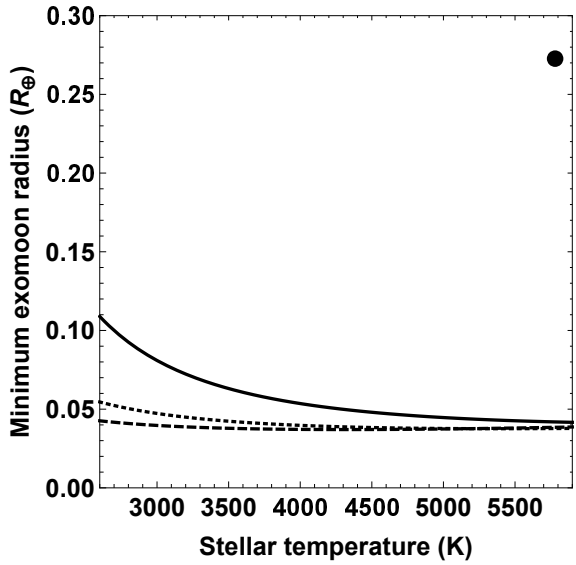} \\
\caption{The minimum moon radius (in $R_\oplus$) necessary for facilitating photosynthesis on the nightside of a tidally locked planet, as a function of the stellar temperature (in K). The unbroken, dotted and dashed curves correspond to the limits for conventional oxygenic photosynthesis (PSI and PSII), $3$-photon and $4$-photon oxygenic photosynthesis schemes, respectively; the associated maximum wavelengths are $\sim 0.7$ $\mu$m, $\sim 1.05$ $\mu$m and $\sim 1.4$ $\mu$m, respectively, as seen from (\ref{Nphdef}). The black dot corresponds to the parameters for Earth's moon and is shown for reference.}
\label{FigMSizev2}
\end{figure}

As we have seen in the preceding paragraph, it is conceivable that near-IR photons might be employed by photoautotrophs deriving their energy from K- and M-dwarfs. It should also be recalled that the wavelength of photons does not directly influence the oxidation of water. Instead, it is the redox potential of the reaction centre in PSII that dictates whether water oxidation is feasible or not; the corresponding redox potential is estimated to be $\sim 1.26$ V \citep{RGND}. In principle, by chaining a number of photosystems together, it is theoretically possible to use photons of longer wavelengths to achieve the oxidation of water and the synthesis of reduced carbon compounds \citep{HB60,HR83,KST07}. However, in doing so, it is important to appreciate that other consequences such as lowered quantum yield may arise as a result. 

Hence, a $\mathcal{N}$-photosystem series utilizing wavelengths up to $\lambda_\mathrm{max}$ can supply the same energy input as the two photosystems (PSI and PSII) of oxygenic photosynthesis, where the relationship between $\lambda_\mathrm{max}$ and $\mathcal{N}$ is given by \citep{WoRa02,KST07}:
\begin{equation}\label{Nphdef}
    \mathcal{N} \approx 2\left(\frac{\lambda_\mathrm{max}}{0.7 \mathrm{\mu m}}\right).
\end{equation}
For the $\mathcal{N}$-photosystem series, the minimum photon flux must be adjusted from $\Phi_c$ to $\left(\mathcal{N}/2\right) \Phi_c$ \citep{WoRa02}. We can repeat the same calculation in Sec. \ref{SecMath} with the modified flux and the adjusted upper wavelength limit. By doing so, we find that the analogue of (\ref{InEqu}) is
\begin{equation}\label{InEquv2}
\left(\frac{A_P}{0.2}\right)\left(\frac{R_P}{R_\oplus}\right)^2 \left(\frac{d}{60\,R_\oplus}\right)^{-2} \left(\frac{T_\star}{T_\odot}\right)^{-1} \mathcal{G}(T_\star) \gtrsim 0.5,
\end{equation}
where the new function $\mathcal{G}(T_\star)$ is defined as
\begin{equation}
\mathcal{G}(T_\star) \approx \frac{2}{\mathcal{N} }\int_{2x_1(T_\star)/ \mathcal{N}}^{x_2(T_\star)} \frac{x'^2\,dx'}{\exp\left(x'\right) - 1}.    
\end{equation}

It is possible to use the above equation to obtain the analogues of the results from Sec. \ref{SecPPM}. Obtaining the appropriate plots is straightforward, and our basic qualitative conclusions are not much affected, owing to which we provide only one example here. For $\mathcal{N} = 3$ and $\mathcal{N} = 4$, the equivalent of Fig. \ref{FigMSize} is plotted in Fig. \ref{FigMSizev2}. The chief differences between higher-order photosystem schemes and conventional oxygenic photosynthesis, with its PSI and PSII, are twoofold. First, for $\mathcal{N} = 3$ and $\mathcal{N} = 4$, we see that the dependence of the moon size on the temperature is weak. Second, we find that the minimum moon size is lowered by a factor of $\lesssim 2$, implying that it must be only $\sim 20\%$ the size of Earth's moon. Thus, by linking a higher number of photosystems, even moons slightly larger than Enceladus (whose radius is $\sim 0.04\,R_\oplus$) might permit oxygenic photosynthesis to function on the planet's nightside under ideal circumstances. 

Now, let us turn our attention to variants of photosynthesis beyond those found on Earth. This subject has received comparatively little attention because of the lack of direct empirical evidence. We focus on a single example for the sake of simplicity, namely, ``hydrogenic photosynthesis''. Studies of exoplanets indicate that many of them may possess substantial hydrogen-helium atmospheres \citep{BRB13,VH17}. On such worlds, \citet{BSZ14} analyzed the prospects for hydrogenic photosynthesis, whose net reaction is given by
\begin{equation}\label{HydroPho}
    \mathrm{CH_4} + \mathrm{H_2O} \rightarrow \mathrm{CH_2O} + 2\mathrm{H_2},
\end{equation}
and it is more instructive to break it down into half-reactions as follows:
\begin{eqnarray}
&& \mathrm{CH_4} + \mathrm{H_2O} \rightarrow \mathrm{CH_2O} + 4\mathrm{H^+} + 4\mathrm{e^-} \nonumber \\
&& \,\, 4\mathrm{H^+} + 4\mathrm{e^-} \rightarrow 2\mathrm{H_2}.
\end{eqnarray}
\citet{BSZ14} proposed that hydrogenic photosynthesis was more advantageous than oxygenic photosynthesis on worlds with hydrogen-dominated atmospheres because the energetic costs in synthesizing a given quantity of biomass are nearly an order of magnitude smaller relative to oxygenic photosynthesis, and the longest wavelength that permits this variant of photosynthesis is $\sim 1.5 \mu$m; in contrast, for conventional photosynthesis the maximum wavelength is around $750$ nm \citep{NMS18}. If we take the latter factor into account and presume that the minimum photon flux for hydrogenic photosynthesis is comparable to $\Phi_c$, we find that (\ref{InEqu}) is transformed into
\begin{equation}\label{InEquHP}
\left(\frac{A_P}{0.2}\right)\left(\frac{R_P}{R_\oplus}\right)^2 \left(\frac{d}{60\,R_\oplus}\right)^{-2} \left(\frac{T_\star}{T_\odot}\right)^{-1} \mathcal{K}(T_\star) \gtrsim 0.5,
\end{equation}
where the new function $\mathcal{G}(T_\star)$ is defined as
\begin{equation}
\mathcal{K}(T_\star) \approx \int_{x_1(T_\star)/2}^{x_2(T_\star)} \frac{x'^2\,dx'}{\exp\left(x'\right) - 1}.    
\end{equation}
We can repeat the analysis in Sec. \ref{SecPPM} using the above two formulae, but we shall not address this topic further as the calculations are fairly straightforward. 

\subsection{Other modes of carbon fixation}\label{SSecCFix}
Hitherto, we have tackled the conditions for photoautotrophy on the nightside of planets and moons. However, even in the case of worlds with permanent nightside that do not receive sufficient photon fluxes, it is crucial to recognize that such worlds might still host fairly diverse biospheres. The primary reason is that photosynthesis does \emph{not} represent the only route to carbon fixation, i.e., the biosynthesis of organic carbon compounds. To put it differently, there are a number of other carbon fixation pathways that can function in the absence of light.

It is instructive to begin by considering the Earth as an example. Most of the biomass on Earth is contributed by photoautotrophs. In particular, land plants (\emph{Embryophyta}) are believed to make up more than $80\%$ of Earth's total biomass \citep{BPM18}. Yet, the contribution of microbes dwelling in deep subsurface habitats is by no means minimal \citep{CH13}. It has been estimated that the majority of Earth's prokaryotes ($>80\%$ by weight) dwell in such environments and make up $\sim 13\%$ of the total biomass \citep{BPM18}. Naturally, not all of these microbes are autotrophs, but it is reasonable to presume that most of them do not rely on phototrophy as these habitats do not have access to sufficient fluxes of PAR photons.

Recent estimates indicate that $>90\%$ of carbon fixation per year by plants, algae and other microbes occurs via the Calvin-Benson-Bassham (CBB) cycle \citep{SS16}, which is also referred to as the reductive pentose phosphate cycle \citep{Berg11}. Aside from the CBB cycle, five other major pathways for carbon fixation have evolved on Earth \citep{fuchs11}. Contemporary studies indicate that they are non-negligible contributors to carbon fixation in Earth's oceans \citep{HS11}. Of these, four of them are cyclic acetyl-CoA–succinyl-CoA pathways that exhibit structural similarities \citep{BFN12}; here, note that CoA signifies coenzyme A. The outlier, and the sixth avenue for carbon fixation, is the reductive acetyl-CoA pathway (also called the Wood-Ljungdahl pathway) - which entails the fixation of two CO$_2$ molecules and leads to the formation of acetyl-CoA - because of its non-cyclic nature \citep{RP08,Berg11}. Aside from the six naturally occurring routes, a synthetic pathway for carbon fixation involving crotonyl-coenzyme A, ethylmalonyl-CoA and hydroxybutyryl-CoA was demonstrated \emph{in vitro} \citep{SS16}.

Despite the dissimilarities among the five pathways aside from the CBB cycle, one of the most striking universal aspects is the central role played by acetyl-coenzyme A (acetyl-CoA). The importance of acetyl-CoA extends beyond its role in carbon fixation pathways because it also regulates mitosis and autophagy, and maintains the balance between cellular anabolism and catabolism \citep{PGB15}. Several hypotheses have, therefore, posited that acetyl-CoA was an essential component of the first metabolic pathway that evolved on Earth \citep{MR06,PGB15}. Of these five networks, the two most important are the reverse tricarboxylic acid (rTCA) cycle and the Wood-Ljungdahl pathway. A combination of physiological, genomic and bioenergetic arguments have been marshalled \citep{SM16,WPX18,NCI18} in conjunction with promising laboratory experiments \citep{MVC17,VMC18,MVM19} to suggest these pathways were the first to emerge on Earth; in fact, certain proposals hypothesize that a hybrid of these two networks might have constituted the ancestral metabolic pathway \citep{BS12,CJV17}.

If, for the sake of argument, we suppose that chemoautotrophy - most likely the rTCA cycle, the Wood-Ljungdahl pathway, or some combination thereof - arose first on other habitable worlds, there still remains the question of how photosynthesis subsequently evolved. With regards to this issue, an important point to note is that many components of the photosynthetic apparatus were probably ported over from chemoautotrophs, with notable examples including: (i) iron-sulfur proteins, (ii) reduced ferredoxins and quinones, and (iii) oxidized electron carriers (e.g., cytochromes and cupredoxins). Hence, it is plausible that (an)oxygenic photosynthesis evolved from chemoautotrophy \citep{SVA13,BG15}. Moreover, RuBisCO exhibits close similarities to other proteins, such as 2,3-diketo-5-methylthiopentyl-1-phosphate enolase, and may have originated from a protein facilitating sulfur metabolism \citep{BG15}.

A number of hypotheses have been put forth to explain how, why and where photosynthesis first arose and the connection to prior carbon fixation pathways. \citet{Nis95} suggested that photosynthesis evolved from phototaxis, with light from hydrothermal vents providing the selective force. \citet{MBB17} proposed that photosynthesis arose to bypass the necessity of flavin-based electron bifurcation to yield reduced ferredoxin utilized in carbon fixation by chemoautotrophs. \citet{MBB17} also conjectured that the high fluxes of ultraviolet (UV) radiation at the surface (see \citealt{CSF07}) hindered the evolution of photosynthesis, and that it emerged instead in the low-intensity IR-dominated regions at hydrothermal vents with Zn-tetrapyrroles constituting the first photopigments. It should, however, be recognized that a number of UV screens potentially existed on early Earth ranging from hazes to biomolecules \citep{Man19}, which could have permitted the evolution of photoautotrophy at the surface.

However, when we consider the permanent nightside of tidally locked planets, the reflected light from a moon is the primary source of radiation. As we have seen earlier, this intensity is orders of magnitude lower than the photon flux incident on Earth. Hence, the aforementioned issue arising from high UV radiation is not applicable. Thus, it seems equally feasible that photosynthesis could arise from prior pathways either on the surface (due to the low-intensity radiation) or near black smokers; note that photoautotrophic green sulfur bacteria (\emph{Chlorobiaceae}) have been detected in the latter environment \citep{BOL05,RaD13}.

\section{Conclusion}\label{SecConc}
The conventional version of photosynthesis experienced on Earth occurs during the day via PAR received directly from the Sun. However, as noted in \citet{RC06} and \citet{CRKL09}, a number of other situations are also feasible for photosynthesis in principle. We have carried out a quantitative analysis of these alternatives for stars, planets and moons of different types.

As tidally locked exoplanets have a permanent nightside, photosynthesis is not conventionally feasible in this hemisphere. However, if the planet has a fairly large moon, the reflected light during the full moon might be capable of powering photosynthesis on the nightside. If viable, photosynthesis would operate with a periodicity equal to the orbital period of the exomoon. By computing the flux of PAR incident on the planet during full moon, we found that the moon must be $\gtrsim 10\%$ the size of the Earth if its albedo is similar to the Earth's moon. Based on dynamical considerations, we argued that ``photosynthesis by moonlight'' is relatively unlikely for planets around M-dwarfs as they have a low likelihood of hosting large, long-lived moons. 

One can reverse this situation and conceive an Earth-like habitable moon orbiting a gas giant that is situated within the habitable zone of a main-sequence star. During the night, reflected light from the planet can illuminate the habitable exomoon and thereby power photosynthesis. We showed that there are regions of parameter space for the planet-moon separation where the exomoon can have a habitable climate while also receiving enough PAR reflected from the planet. However, because of tidal heating and orbital stability, habitable exomoons are unlikely to exist around late-type M-dwarfs (with $M_\star \lesssim 0.2 M_\odot$).

Although we have determined that a photosynthesis-based biosphere is permitted for a wide range of stars and planet-moon separations, the NPP of the corresponding biosphere might be much lower compared to the Earth's biosphere. In fact, if we assume that the biosphere is photon-limited, i.e., restricted by PAR flux, the NPP on the nightside of a tidally locked exoplanet due to reflected moonlight will be $\sim 5$ orders of magnitude smaller than the Earth's NPP. Of course, one should recognize that other physical and chemical constraints also govern the NPP such as the access to nutrients, water, and reactants as well as the ambient temperature. Furthermore, as noted in Sec. \ref{SSecCFix}, other carbon fixation pathways can contribute toward the NPP and the sustenance of biospheres even in the absence of light.

In summary, we have investigated the constraints on photosynthesis via reflected light from one object incident on another object in a planet-moon system situated in the habitable zone of the host star. Our analysis indicates that this variant of photosynthesis may be feasible, although by no means guaranteed, provided that $M_\star \gtrsim 0.2 M_\odot$.\footnote{Planets and moons orbiting late-type M-dwarfs could experience additional habitability issues such as atmospheric erosion by strong stellar winds, high ionizing radiation doses, and insufficient photon fluxes for prebiotic chemistry \citep{DLMC,DJL18,Ling18,LL18,ML18,Man19}.}

\section*{Acknowledgments}
We thank the reviewer for the valuable comments regarding the paper. This work was supported in part by the Breakthrough Prize Foundation, Harvard University's Faculty of Arts and Sciences, and the Institute for Theory and Computation (ITC) at Harvard University.


\end{document}